\documentclass[aps,prl,twocolumn,superscriptaddress,showpacs]{revtex4-1}

\usepackage{graphicx,color}
\usepackage{bm}
 
\begin{document}


\title{Crossover from skin-mode to proton-neutron-mode
in $E1$ excitations of neutron-rich nuclei}


\author{H.~Nakada}
\email[E-mail:\,\,]{nakada@faculty.chiba-u.jp}
\affiliation{Department of Physics, Graduate School of Science,
 Chiba University\\
Yayoi-cho 1-33, Inage, Chiba 263-8522, Japan}

\author{T.~Inakura}
\affiliation{Department of Physics, Graduate School of Science,
 Chiba University\\
Yayoi-cho 1-33, Inage, Chiba 263-8522, Japan}
\affiliation{Nishina Center for Accelerator-Based Science, RIKEN\\
Hirosawa 2-1, Wako, Saitama 351-0198, Japan}

\author{H.~Sawai\,}
\affiliation{Department of Physics, Graduate School of Science,
 Chiba University\\
Yayoi-cho 1-33, Inage, Chiba 263-8522, Japan}


\date{\today}

\begin{abstract}
Character of the low-energy $E1$ excitations is investigated
by analyzing transition densities
obtained from the RPA calculations in the doubly-magic nuclei.
We propose a decomposition method of the $E1$ excitations
into the $pn$ mode (\textit{i.e.} oscillation between protons and neutrons)
and the skin mode (\textit{i.e.}
oscillation of the neutron skin against the inner core)
via the transition densities,
by which their mixing is handled in a straightforward manner.
Crossover behavior of the $E1$ excitations is found,
from the skin mode at low energy to the $pn$ mode at higher energy.
The ratio of the skin-mode strength to the full strength
turns out to be insensitive to nuclide and to effective interactions
in the energy region of the crossover.
Depending on the excitation energy,
the observed low-energy $E1$ excitations are not necessarily dominated
by the skin mode, as exemplified for $^{90}$Zr.
\end{abstract}


\maketitle



\textit{Introduction.}
While it has been known that the $E1$ excitations in atomic nuclei
are predominantly concentrated in the giant dipole resonance (GDR),
recent experiments have disclosed
that sizable $E1$ strengths are present at lower energies
in a number of $N>Z$ nuclei;
in nuclei far off the $\beta$ stability
such as $^{68}$Ni~\cite{Wieland09} and $^{130,132}$Sn~\cite{Adrich05},
and in medium- and heavy-mass nuclei along the $\beta$ stability line,
$^{90}$Zr~\cite{Schwengner,Iwamoto},
the Pb isotopes~\cite{Chapuran,Ryezayeva,Endres00,Tamii,Poltoratska}
and the $N=82$ isotones~\cite{Herzberg,Zilges,Volz,Savran06,Savran08,
Endres09,Endres10,Tonchev}.
These low-energy $E1$ strengths are often called
pygmy dipole resonance (PDR),
though their physical content yet remains obscure.
The GDR is a typical collective excitation of nuclei,
in which the constituent protons oscillate against neutrons
($pn$ mode), keeping the center-of-mass (c.m.) at rest.
A possible interpretation of the PDR is
an oscillation of the neutron skin against the inner core (skin mode),
as suggested in pioneering studies
with macroscopic models~\cite{mohan71,suzuki90,isacker92}.
However, recent calculations using the random-phase approximation (RPA)
do and do not support the skin-mode picture of the observed PDRs,
depending on nuclides as well as on effective interactions
or energy density functionals.
A systematic calculation~\cite{INY09} shows that
the low-energy $E1$ strengths well correlate to the neutron skin thickness
in some regions of nuclides,
but not in other regions.
The low-energy $E1$ strengths are quite relevant
to reaction rates in nucleosynthesis.
Moreover, the low-energy $E1$ strengths could correlate
to density-dependence of the nuclear symmetry energy,
and therefore to structure of neutron stars.
It is important to comprehend energy- and nucleus-dependence
of the low-energy $E1$ excitations,
to which inspection of its character should be crucial.

The $E1$ transitions have isovector (IV) character
in the long wavelength limit.
In the pure $pn$ mode the transition densities should be IV
at any position.
However, the transition densities at each position
may have isoscalar (IS) character in the skin mode,
although the IS component should vanish after integration
(see Eq.~(\ref{eq:int-sptrd}) given later).
In Refs.~\cite{Savran06,Endres09,Endres10},
it has been shown experimentally
that low-energy $1^-$ states in $^{138}$Ba, $^{140}$Ce and $^{124}$Sn
can be excited by $(\alpha,\alpha^\prime\gamma)$
as well as by $(\gamma,\gamma^\prime)$,
in contrast to higher-energy states excited only via $(\gamma,\gamma^\prime)$.
This seems to support the skin-mode interpretation
of the low-energy $E1$ strengths in these nuclei.
The isospin character of the low-energy transitions has been investigated
for $^{140}$Ce~\cite{Paar09}, $^{68}$Ni, $^{132}$Sn
and $^{208}$Pb~\cite{Roca-Maza12,Vretenar12} by the RPA calculations.
It has been pointed out in those studies
that character of the transitions varies as energy increases.
Similar results have been reported from calculations
in the quasiparticle-phonon model~\cite{Tsoneva08}.
However, in these experimental and theoretical studies,
character of the $E1$ transitions does not necessarily switch
from the skin mode to the $pn$ mode sharply.
It is hence desired
to investigate character of the low-energy $E1$ transitions
with taking admixture of the two modes into consideration.
Of particular interest is how the degree of the mixing varies
for increasing energy and/or mass number.
Since the transition operators to $1^-$ states
have different radial-dependence between the IS and the IV channels
so as not to contain the c.m. motion,
the IS and IV transition strengths cannot directly be compared.
We should also notice that the oscillation of the neutron skin
against the inner core should provide transition density
dominated by neutrons outside the core,
which has IV as much as IS characters.
It is not easy to draw quantitative information of the mixing
between the $pn$ and the skin modes
only from the isospin character of the transition rates,
although it is certainly useful for qualitative arguments.

In this article
we investigate character of the low-energy $E1$ excitations
via the RPA calculations on top of the Hartree-Fock (HF) wave functions.
Calculations have been implemented for a number of doubly magic nuclei
with various non-relativistic effective interactions,
the Skyrme (SkM$^\ast$, SkI2)~\cite{SkMs,SkI},
the Gogny (D1S, D1M)~\cite{D1S,D1M}
and the M3Y-type~(M3Y-P7)~\cite{Nak03,Nak12} interactions,
by using the numerical methods of Refs.~\cite{Ina06,NIY07,NMYM09}.
Note that in these calculations
the spurious c.m. motion has been separated out to good precision,
both for the Skyrme and the finite-range interactions.
We then introduce a new method of decomposing the low-energy $E1$ strengths
into the $pn$ and the skin modes
from the calculated transition densities,
by which mixing of the two modes is taken into account
in a straightforward manner.
Via this decomposition we shall examine
energy- and nucleus-dependence of character of the $E1$ transitions.
Since interaction-dependence does not influence essential points of arguments,
we shall mainly show results of the SkI2 interaction below,
with which observed PDRs in several nuclei are reproduced
relatively well~\cite{Car10}.

\textit{Transition density.}
The $E1$ transition operator is expressed as
\begin{equation}
 \mathcal{O}^{(E1)}
 = \frac{N}{A}\sum_{i\in p} r_i Y^{(1)}(\hat{\mathbf{r}}_i)
  -\frac{Z}{A}\sum_{i\in n} r_i Y^{(1)}(\hat{\mathbf{r}}_i)\,,
  \label{eq:op1}
\end{equation}
after the c.m. correction.
Here $i$ is the index of nucleons
and $i\in p$ ($i\in n$) indicates that the sum runs over protons (neutrons).
The $E1$ strength from the ground state $|0\rangle$
to the $1^-$ state $|\alpha\rangle$ in a even-even nucleus is
$B(E1;\alpha) = \Big|\langle\alpha|\mathcal{O}^{(E1)}|0\rangle\Big|^2$.
As the $1p$-$1h$ excitations,
we calculate the transition amplitude $\langle\alpha|\mathcal{O}^{(E1)}|0\rangle$
within the HF+RPA.
By smearing out with Lorentzian,
the strength function is given by
\begin{eqnarray}
 S^{(E1)}(\omega) &=& \frac{\gamma}{\pi}\sum_\alpha
  \bigg[\frac{1}{(\omega-\omega_\alpha)^2+\gamma^2}\nonumber\\
  &&\quad - \frac{1}{(\omega+\omega_\alpha)^2+\gamma^2}\bigg]\,
  \Big|\langle\alpha|\mathcal{O}^{(E1)}|0\rangle\Big|^2\,.
 \label{eq:S(E)}
\end{eqnarray}
We adopt $2\gamma=1\,\mathrm{MeV}$ throughout this paper.

We next define the proton and neutron transition density by
\begin{equation}
 r^2\delta\rho_{\tau_z}^{(\lambda=1)}(r;\alpha)
  = \left\{\begin{array}{r}
 {\displaystyle\langle \alpha|\sum_{i\in p} \delta(r-r_i)\,
  r_i Y^{(1)}(\hat{\mathbf{r}}_i)|0\rangle} \\
 (\mbox{for~}\tau_z=p) \\
 {\displaystyle\langle \alpha|\sum_{i\in n} \delta(r-r_i)\,
  r_i Y^{(1)}(\hat{\mathbf{r}}_i)|0\rangle} \\
 (\mbox{for~}\tau_z=n) \end{array}
 \right.\,.
\end{equation}
Associated with $\mathcal{O}^{(E1)}$ in Eq.~(\ref{eq:op1}),
the $E1$ transition density is defined as
\begin{equation}
 \delta\rho^{(E1)}(r;\alpha) = \frac{N}{A}\delta\rho_p^{(\lambda=1)}(r;\alpha)
  -\frac{Z}{A}\delta\rho_n^{(\lambda=1)}(r;\alpha)\,,
 \label{eq:trdns1}
\end{equation}
which yields $\langle\alpha|\mathcal{O}^{(E1)}|0\rangle
= \int r^2dr\,\delta\rho^{(E1)}(r;\alpha)$.
It is noticed that, because the spurious c.m. motion is well separated,
we have
\begin{equation}
 \int r^2 dr\,\big[\delta\rho_p^{(\lambda=1)}(r;\alpha)
 + \delta\rho_n^{(\lambda=1)}(r;\alpha)\big]
  = 0\,.
\label{eq:int-sptrd}
\end{equation}

The transition density $\delta\rho_{\tau_z}^{(\lambda=1)}(r)$
is useful to investigate character of an individual transition.
In the $pn$ mode we have $\delta\rho_p^{(\lambda=1)}(r)
+\delta\rho_n^{(\lambda=1)}(r)\approx 0$ at any $r$.
In contrast, in the skin mode $\delta\rho_p^{(\lambda=1)}(r)$
and $\delta\rho_n^{(\lambda=1)}(r)$ have equal phases
up to around the nuclear surface,
and $\delta\rho_n^{(\lambda=1)}(r)$ at larger $r$
has the opposite phase to them,
so that Eq.~(\ref{eq:int-sptrd}) should be fulfilled.
This indicates that $\delta\rho^{(E1)}(r)$ is small in the interior
because $\delta\rho_p^{(\lambda=1)}(r)$ and $\delta\rho_n^{(\lambda=1)}(r)$
tend to cancel each other,
while $\delta\rho^{(E1)}(r)\,(\approx(-Z/A)\,\delta\rho_n^{(\lambda=1)}(r))$
at large $r$ mainly contributes to the $E1$ strength.

\begin{figure}
\includegraphics[scale=0.4]{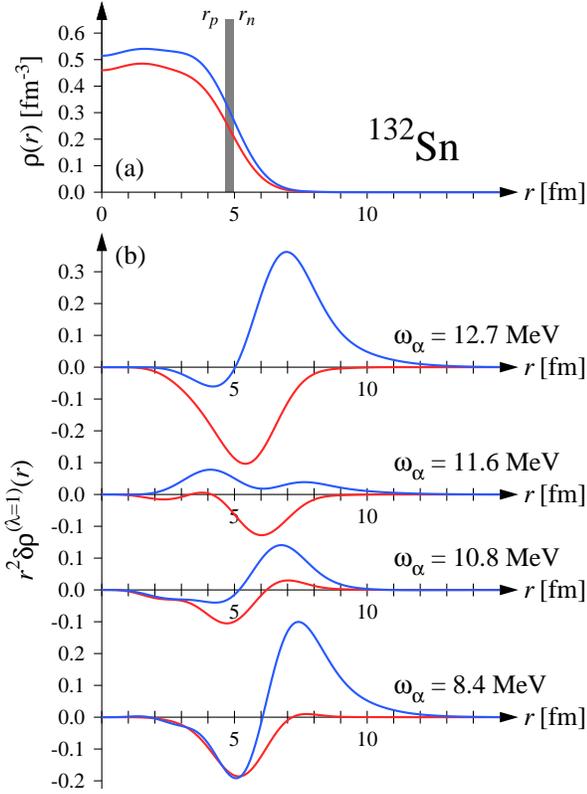}
\caption{(Color online)
(a) Density distribution $\rho_p(r)$ (red line) and $\rho_n(r)$ (blue line)
 of the ground state in $^{132}$Sn by the HF calculation with SkI2.
 The thick gray bar is bounded by the proton and neutron rms radii
 ($r_p$ and $r_n$), presenting their difference.
(b) Transition densities $r^2\delta\rho_p^{(\lambda=1)}(r;\alpha)$ (red line)
 and $r^2\delta\rho_n^{(\lambda=1)}(r;\alpha)$ (blue line)
 to several $1^-$ states in $^{132}$Sn by the HF+RPA calculation with SkI2.
 The overall phases are taken so as for the highest peak
 of $r^2\delta\rho_n^{(\lambda=1)}(r;\alpha)$ to be positive, for each $\alpha$.
 \label{fig:trdns-Sn132}}
\end{figure}

\begin{figure}
\includegraphics[scale=0.4]{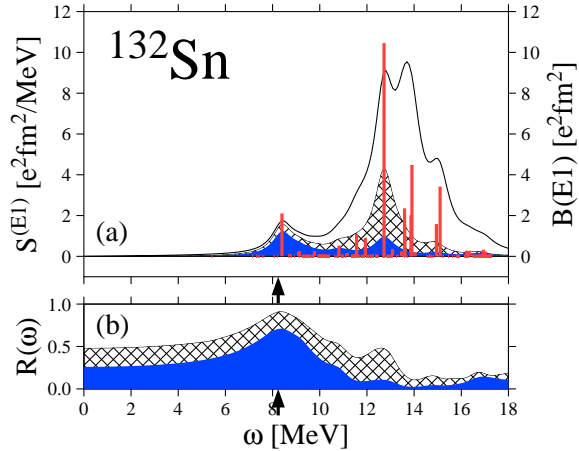}
\caption{(Color online)
(a) Decomposed $E1$ strength functions
 and $B(E1;\alpha)$ in $^{132}$Sn.
 Solid line shows
 $S^{(E1)}(\omega)\,\big(=S^{(E1)}_{pn}(\omega)+S^{(E1)}_\mathrm{skin}(\omega)
 +S^{(E1)}_\mathrm{intf}(\omega)\big)$,
 while blue-shaded and hatched areas present
 $S^{(E1)}_\mathrm{skin}(\omega)$ and $S^{(E1)}_\mathrm{intf}(\omega)$, respectively.
 Red vertical bars display $B(E1;\alpha)$ without smoothing.
 Black arrow attached to the horizontal axis indicates
 the neutron threshold in the HF calculation.
(b) $R_\mathrm{skin}(\omega)$ and $R_\mathrm{intf}(\omega)$.
 Conventions correspond to those of the strength functions in (a).
 The SkI2 interaction is employed.
 \label{fig:Sn132E1-decomp}}
\end{figure}

In Fig.~\ref{fig:trdns-Sn132}(b)
$r^2\delta\rho_{\tau_z}^{(\lambda=1)}(r;\alpha)$
is depicted for several excited states $\alpha$ in $^{132}$Sn
(also see Fig.~\ref{fig:Sn132E1-decomp} for the strength function).
For reference, we show the proton and neutron density distributions
in the HF ground state $\rho_{\tau_z}(r)$ in Fig.~\ref{fig:trdns-Sn132}(a),
along with the proton and neutron rms radii.
The excitation to the $\omega_\alpha=8.4\,\mathrm{MeV}$ state
has typical transition densities of the skin mode,
while the excitation to $\omega_\alpha=12.7\,\mathrm{MeV}$
is dominated by the $pn$ mode,
although we may view skin-mode-like behavior
in $r\lesssim 5\,\mathrm{fm}$ and in the displacement of the peaks
between $r^2\delta\rho_p^{(\lambda=1)}(r)$ and $r^2\delta\rho_n^{(\lambda=1)}(r)$.
However, at energies in-between it is difficult
to classify the excitations into either of the $pn$ or the skin mode.
As the two modes seem to mix at these intermediate energies,
the transition densities gradually change as $\omega$ increases.
In this respect character change of the $E1$ strengths is not like
a phase transition from the skin mode to the $pn$ mode,
should rather be regarded as \textit{crossover} between the two modes.

\textit{Decomposition of $E1$ strengths.}
It is reasonable to assume that $E1$ strengths are dominated
by the two components,
one corresponds to the skin mode while the other to the $pn$ mode,
up to around the GDR peak,
though they can mix in individual excitations.
In order to view the gradual change of the $E1$ character closely,
we separate the $E1$ transition density to each $|\alpha\rangle$
into the $pn$- and the skin-mode components.
Although it is not obvious how the $E1$ strengths can be decomposed
into these two ingredients,
we here introduce a method which works practically well,
particularly in handling their mixing.
As argued above,
while the $pn$-mode transitions hold IV character
with $\delta\rho_p^{(\lambda=1)}(r)$
and $\delta\rho_n^{(\lambda=1)}(r)$ having opposite phases,
in the skin mode the transition densities have IS character
in a certain spatial region,
basically the interior of the nucleus,
and are dominated by neutrons in the remaining region.
We therefore separate the excitation to each $|\alpha\rangle$
into the $pn$- and the skin-mode components in a position-dependent manner
by the following criterion:
\begin{eqnarray} &&\left.\begin{array}{l}
 \delta\rho_{pn}^{(E1)}(r;\alpha) = 0 \\
 \delta\rho_\mathrm{skin}^{(E1)}(r;\alpha) = \delta\rho^{(E1)}(r;\alpha)
 \end{array}\right\}\nonumber\\
 &&\qquad\mbox{for $r$ satisfying }
 {\displaystyle\frac{\delta\rho_p^{(\lambda=1)}(r;\alpha)}
 {\delta\rho_n^{(\lambda=1)}(r;\alpha)} > -\lambda_s}\,,\nonumber\\
&&\left.\begin{array}{l}
 \delta\rho_{pn}^{(E1)}(r;\alpha) = \delta\rho^{(E1)}(r;\alpha) \\
 \delta\rho_\mathrm{skin}^{(E1)}(r;\alpha) = 0 \end{array}\right\}\nonumber\\
 &&\qquad\mbox{otherwise}\,. \label{eq:criterion}\end{eqnarray}
Obviously $\delta\rho^{(E1)}(r;\alpha)
= \delta\rho_{pn}^{(E1)}(r;\alpha)+\delta\rho_\mathrm{skin}^{(E1)}(r;\alpha)$.
To pick up components with $\delta\rho_p^{(\lambda=1)}(r)\approx 0$
as the skin mode,
it is appropriate to take $0<\lambda_s\ll 1$.
For the $12.7\,\mathrm{MeV}$ excitation of Fig.~\ref{fig:trdns-Sn132},
the transition density at $r\lesssim 5\,\mathrm{fm}$
and $9\lesssim r\lesssim 13\,\mathrm{fm}$ is classified as
$\delta\rho_\mathrm{skin}^{(E1)}(r;\alpha)$,
while the dominant part of $5\lesssim r\lesssim 9\,\mathrm{fm}$
as $\delta\rho_{pn}^{(E1)}(r;\alpha)$.
Whereas it may be questioned
whether the criterion of Eq.~(\ref{eq:criterion}) is appropriate
for very weak $E1$ excitations or $E1$ excitations at high energy,
it will be useful in investigating energy-dependent character
of the $E1$ strengths below the GDR peak.
Since the IS-like transition density does not significantly contribute
to the $E1$ strength,
the condition of Eq.~(\ref{eq:criterion}) indicates
dominance of the neutron excitation for $\delta\rho_\mathrm{skin}^{(E1)}(r)$,
as long as transitions at relatively low energy are concerned.
This separation of the spatial region according to the transition densities
provides the way to decompose the $E1$ transition matrix element,
\begin{eqnarray}
&& \langle\alpha|\mathcal{O}^{(E1)}|0\rangle
 =\langle\alpha|\mathcal{O}^{(E1)}|0\rangle_{pn}
 +\langle\alpha|\mathcal{O}^{(E1)}|0\rangle_\mathrm{skin}\,; \nonumber\\
&&\quad \left\{\begin{array}{ll}
  {\displaystyle \langle\alpha|\mathcal{O}^{(E1)}|0\rangle_{pn}
   =\int r^2dr\,\delta\rho_{pn}^{(E1)}(r;\alpha)}\\
  {\displaystyle \langle\alpha|\mathcal{O}^{(E1)}|0\rangle_\mathrm{skin}
   =\int r^2dr\,\delta\rho_\mathrm{skin}^{(E1)}(r;\alpha)}\end{array}\right.\,,
\end{eqnarray}
and further to the decomposition of the strength function,
\begin{eqnarray}
&& S^{(E1)}(\omega)=S^{(E1)}_{pn}(\omega)+S^{(E1)}_\mathrm{skin}(\omega)
 +S^{(E1)}_\mathrm{intf}(\omega)\,; \nonumber\\
&&\quad \left\{\begin{array}{ll} S^{(E1)}_{pn}(\omega)
   &={\displaystyle \frac{\gamma}{\pi}\sum_\alpha
  \bigg[\frac{1}{(\omega-\omega_\alpha)^2+\gamma^2}}\\
  &\qquad{\displaystyle - \frac{1}{(\omega+\omega_\alpha)^2+\gamma^2}\bigg]}\,
  \Big|\langle\alpha|\mathcal{O}^{(E1)}|0\rangle_{pn}\Big|^2\\
  S^{(E1)}_\mathrm{skin}(\omega)
   &={\displaystyle \frac{\gamma}{\pi}\sum_\alpha
  \bigg[\frac{1}{(\omega-\omega_\alpha)^2+\gamma^2}}\\
  &\qquad{\displaystyle - \frac{1}{(\omega+\omega_\alpha)^2+\gamma^2}\bigg]}\,
  \Big|\langle\alpha|\mathcal{O}^{(E1)}|0\rangle_\mathrm{skin}\Big|^2\\
  S^{(E1)}_\mathrm{intf}(\omega)
   &={\displaystyle \frac{\gamma}{\pi}\sum_\alpha
  \bigg[\frac{1}{(\omega-\omega_\alpha)^2+\gamma^2}}\\
  &\qquad{\displaystyle - \frac{1}{(\omega+\omega_\alpha)^2+\gamma^2}\bigg]}\\
  &\hfill \times 2\,\mathrm{Re}\Big[
    \langle\alpha|\mathcal{O}^{(E1)}|0\rangle_{pn}^\ast\,
    \langle\alpha|\mathcal{O}^{(E1)}|0\rangle_\mathrm{skin}\Big]
 \end{array}\right.\,.
\end{eqnarray}

The results of decomposing the $E1$ strength functions with $\lambda_s=0.05$
are displayed in Fig.~\ref{fig:Sn132E1-decomp}(a),
along with $B(E1)$ to an individual state $|\alpha\rangle$,
for $^{132}$Sn.
By this decomposition,
we view more clearly the crossover from the skin mode to the $pn$ mode
in the $E1$ transition strengths.
We have confirmed that the results do not change much
with $\lambda_s=0.1$ and $\lambda_s=0.2$~\cite{Sawai}.
Moreover, although the decomposition may reflect
character of individual states $|\alpha\rangle$,
the decomposed strength functions do not have strong irregularity,
after smoothing with respect to $\omega$.

We further define the ratios
\begin{equation}
 R_\mathrm{mode}(\omega) = \frac{S^{(E1)}_\mathrm{mode}(\omega)}
  {S^{(E1)}(\omega)}\,;\quad(\mbox{`$\mathrm{mode}$'\,$=$\,`$pn$',
  `$\mathrm{skin}$' or `$\mathrm{intf}$'})\,, \label{eq:def-R}
\end{equation}
which are exemplified for $^{132}$Sn
in Fig.~\ref{fig:Sn132E1-decomp}(b).
Note that $R_{pn}(\omega)=1-R_\mathrm{skin}(\omega)-R_\mathrm{intf}(\omega)$.
These ratios are not sufficiently meaningful
for $\omega$ at which $S^{(E1)}(\omega)$ is quite small,
namely at very low $\omega$ (in absence of sizable $E1$ strengths)
as well as at high $\omega$ beyond GDR.
However, above the energy
at which $R_\mathrm{skin}(\omega)$ starts decreasing
($\omega\gtrsim 8\,\mathrm{MeV}$ in Fig.~\ref{fig:Sn132E1-decomp}),
$R_\mathrm{skin}(\omega)$ keeps decreasing almost monotonically
up to the GDR peak energy, apart from small fluctuations.
This illustrates that the position-dependent decomposition
of the $E1$ strengths of Eq.~(\ref{eq:criterion}) works practically well
at $\omega\lesssim 20\,\mathrm{MeV}$.

\begin{figure}
\includegraphics[scale=0.3]{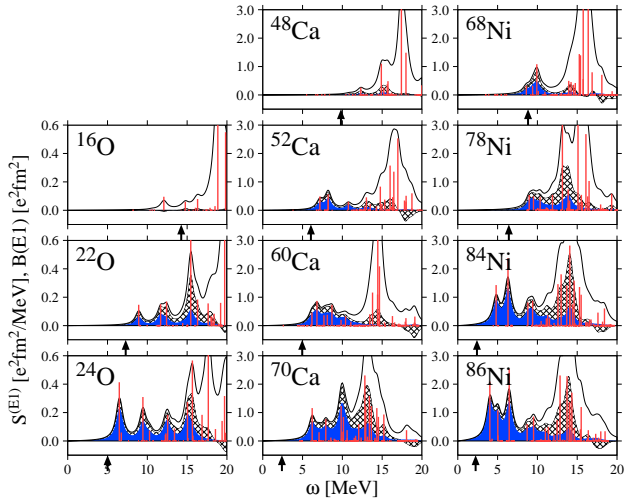}
\caption{(Color online)
 Decomposed $E1$ strength functions and $B(E1;\alpha)$ with SkI2.
 See Fig.~\protect\ref{fig:Sn132E1-decomp} for conventions.
 \label{fig:E1-decomp}}
\end{figure}

\begin{figure}
\includegraphics[scale=0.3]{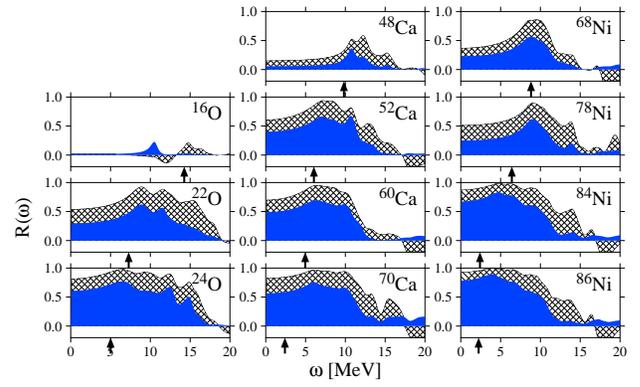}
\caption{(Color online)
 $R_\mathrm{skin}(\omega)$ and $R_\mathrm{intf}(\omega)$ with SkI2.
 See Fig.~\protect\ref{fig:Sn132E1-decomp} for conventions.
 \label{fig:E1-ratio}}
\end{figure}

\begin{figure}
\includegraphics[scale=0.3]{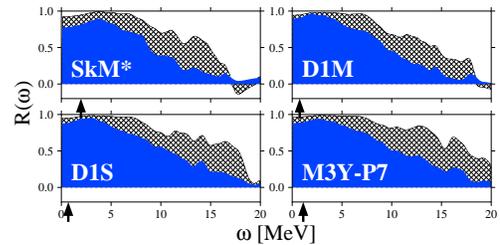}
\caption{(Color online)
 Comparison of $R_\mathrm{skin}(\omega)$ and $R_\mathrm{intf}(\omega)$
 in $^{86}$Ni among various effective interactions.
 See Fig.~\protect\ref{fig:Sn132E1-decomp} for conventions.
 \label{fig:E1-ratio_Ni86}}
\end{figure}

The decomposed strength functions and the ratios
are depicted for many doubly magic (or nearly doubly magic) nuclei
in Figs.~\ref{fig:E1-decomp} and \ref{fig:E1-ratio}.
We confirm that the essential points found in $^{132}$Sn
hold for other nuclei.
In neutron-rich nuclei there is a certain energy region
where $R_\mathrm{skin}(\omega)$ exceeds $0.5$,
although $R_\mathrm{skin}(\omega)$ values are not always quite meaningful
at very low $\omega$ as in $^{132}$Sn.
Let us give special attention to the energy
at which $R_\mathrm{skin}(\omega)$ comes down to $0.5$
(after exceeding $0.5$) in these neutron-rich nuclei.
It is relatively high in the lighter nuclei $^{22,24}$O.
However, the energy at which $R_\mathrm{skin}(\omega)$ decreases down to $0.5$
is $\omega\approx 10\,\mathrm{MeV}$
for all nuclei from $^{52}$Ca to $^{132}$Sn,
with strikingly weak nucleus-dependence beyond the Ca region.
This consequence hardly depends on the effective interactions,
since $R_\mathrm{skin}(\omega)$ in the crossover energy region
is not quite sensitive to the effective interaction,
as illustrated for $^{86}$Ni in Fig.~\ref{fig:E1-ratio_Ni86}.

\textit{PDR in $^{90}$Zr.}
Because of the mixing of the $pn$ mode and the skin mode,
it is not guaranteed that all the observed low-energy $E1$ strengths
should be attributed to the skin mode.
Therefore careful study is needed in interpreting observed PDRs.
We here argue character of the $E1$ strengths in $^{90}$Zr as an example.

\begin{figure}
\includegraphics[scale=0.4]{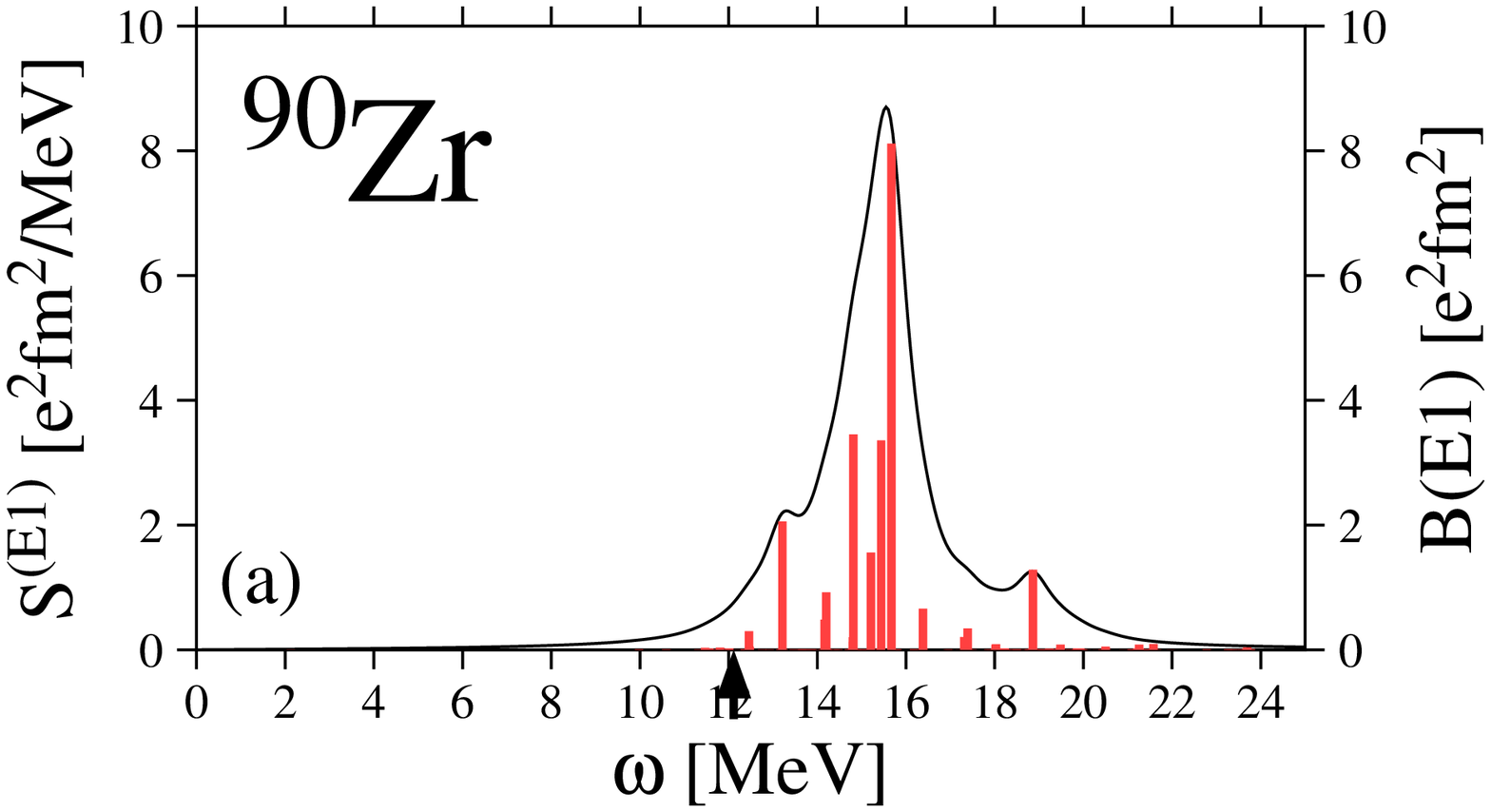}
\includegraphics[scale=0.4]{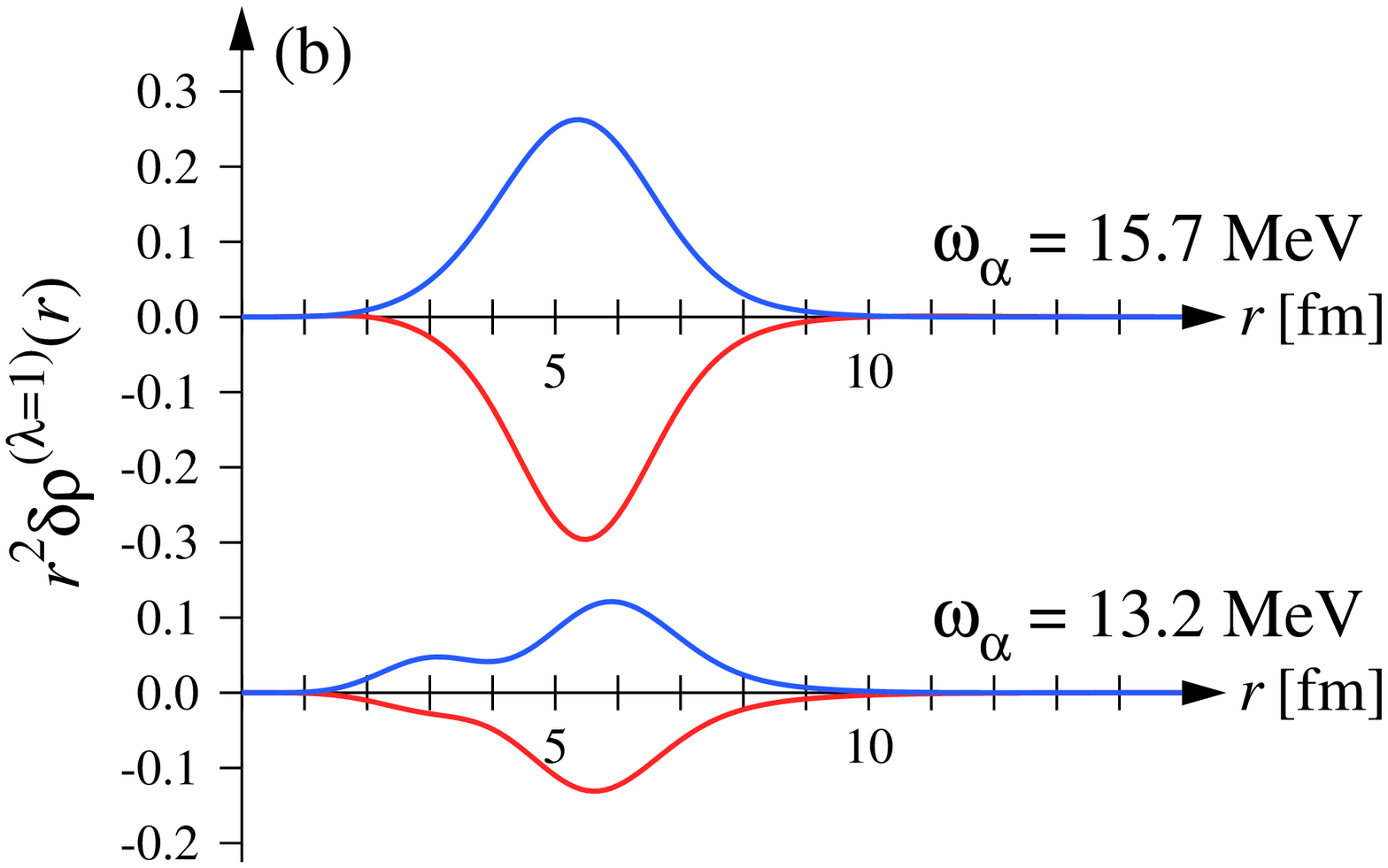}
\caption{(Color online)
(a) $S^{(E1)}(\omega)$, $B(E1;\alpha)$
and (b) $r^2\delta\rho^{(\lambda=1)}(r;\alpha)$ in $^{90}$Zr, with SkI2.
 See Figs.~\ref{fig:trdns-Sn132} and \ref{fig:Sn132E1-decomp}
 for conventions.
 \label{fig:Zr90}}
\end{figure}

The strength function and transition densities with SkI2
are shown in Fig.~\ref{fig:Zr90},
in which a shoulder structure is found at $\omega\approx 13\,\mathrm{MeV}$.
This shoulder state is a good $pn$ mode but with two bumps
in the transition densities.
The other currently used interactions give similar results.

Recent experiments have established a peak at $\omega\approx 9\,\mathrm{MeV}$
in the $E1$ strength distribution of $^{90}$Zr~\cite{Schwengner,Iwamoto}.
Suppose that the calculated strength at $\omega\approx 13\,\mathrm{MeV}$
corresponds to this observed PDR,
the PDR in $^{90}$Zr is not associated with the skin mode,
rather being a result of fragmentation of the $pn$ mode.
Despite the displacement in energy,
it is still suggested that the observed PDR is not highly
dominated by the skin mode,
since mixing of the two modes is expected at $\omega\approx 9\,\mathrm{MeV}$,
according to nucleus-insensitive behavior of $R_\mathrm{skin}(\omega)$
in Fig.~\ref{fig:E1-ratio}.
Experimental information on the transition density will be desired.

\textit{Summary.}
By analyzing the transition densities
obtained from the HF+RPA calculations in the doubly-magic nuclei,
we have investigated character of the low-energy $E1$ excitations.
In a certain energy region
the $pn$ mode (\textit{i.e.} oscillation between protons and neutrons)
and the skin mode (\textit{i.e.} oscillation
of the neutron skin against the inner core)
mix with each other.
We propose a method to decompose the $E1$ excitations
into the $pn$ mode and the skin mode via the transition densities,
which practically works well
and enables us to handle their mixing explicitly.
It has been shown that the $E1$ excitations gradually change their character
as the energy increases.
This behavior may be regarded as crossover transition
from the skin mode at low energy to the $pn$ mode at higher energy.
The skin-mode strength comes down to a half of the full $E1$ strength
at $\omega\approx 10\,\mathrm{MeV}$ from $^{52}$Ca to $^{132}$Sn,
generic for all nuclide and effective interactions
under investigation
as far as sizable $E1$ strengths are present at low energy.

Because of the energy-dependence of the character of the $E1$ excitations,
it is not obvious whether observed PDR in an individual nucleus
is dominated by the skin mode.
We argue the case of $^{90}$Zr as an example,
via the calculated strength function
as well as via the generic energy-dependence of the skin-mode fractions.
We emphasize importance of experimental information
of the transition densities.

This work is financially supported
as Grant-in-Aid for Scientific Research No.~24105008 by MEXT,
and No.~22540266 by JSPS.
Numerical calculations have been performed
on HITAC SR16000s at IMIT in Chiba University,
at IIC in Hokkaido University, at RIIT in Kyushu University,
at YITP in Kyoto University and at ITC in University of Tokyo.


\end{document}